\documentclass[12pt]{iopart}

\usepackage{amssymb,amsfonts,setstack}

\def\be{\begin{equation}}
\def\ee{\end{equation}}
\def\bea{\begin{eqnarray}}
\def\eea{\end{eqnarray}}
\def\be{\begin{equation}}
\def\ee{\end{equation}}

\newcommand\fverb{\setbox\pippobox=\hbox\bgroup\verb}
\newcommand\fverbdo{\egroup\medskip\noindent%
                        \fbox{\unhbox\pippobox}\ }
\newcommand\fverbit{\egroup\item[\fbox{\unhbox\pippobox}]}
\newbox\pippobox

\def\F{\Phi}

\def\m{\mu}

\def\s{\sigma}

\def\p{\partial}

\def\f{\varphi}
\def\a{\alpha}

\def\d{\delta}
\def\l{\lambda}
\def\g{\gamma}
\def\G{\Gamma}

\def\w{\Omega}
\def\ba{\begin{eqnarray}}
\def\ea{\end{eqnarray}}

\def\jh{{\hat{\jmath}}}

\def\np#1#2#3{Nucl. Phys. {\bf{B#1}} (#2) #3}

\begin{document}


\title{Gravitational waves from WZW models}

\author{Giuseppe D'Appollonio}

\address{
LPTHE, Universit\'e Paris VI, 4 pl Jussieu, \\
75252 Paris cedex 05, FRANCE}

\ead{giuseppe@lpthe.jussieu.fr}

\begin{abstract}

A brief review is given of the recent solution of a 
non-compact CFT describing a NS-supported pp-wave background. 
We will first explain  how to compute the three and four-point 
correlators using current algebra techniques, thereby showing 
that some generic features of the non-compact WZW models become
very clear in this simple context. We will then present the 
Penrose limit as a contraction of an $U(1) \times SU(2)_k$ 
WZW model, an approach that could prove useful in order to 
understand holography for pp-wave space-times. We will finally 
comment on the string amplitudes and on the existence of two flat 
space limits. 

\end{abstract}


Plane-fronted gravitational wave backgrounds are exact solutions 
of the classical string equations of motion \cite{gak}, a property
that follows from the existence of a covariantly 
constant null vector in these space-times \cite{hs}. 
They represent a large class of time-dependent 
and possibly singular solutions of string theory that enable the investigation
of some non-trivial aspects of the string dynamics
by quantizing the world-sheet $\s$-model in the light-cone gauge
(for a recent discussion see \cite{prt}).
Whenever the $\s$-model displays a larger symmetry,
such as a current algebra symmetry, we can study
the properties of the corresponding pp-wave backgrounds in 
more detail and in particular we can expect to be able to compute some correlation
functions. The first example of a gravitational wave related to a 
WZW model was discovered by Nappi and Witten \cite{nw}.
They considered a WZW model based on the Heisenberg group $H_4$, 
defined by the following commutation relations: 
\be [P^+,P^-] = -2i \m K \ , \hspace{2.8cm} [J,P^{\pm}] = \mp i \m
P^{\pm} \ . \label{a1}  \ee
Since the group is not semi-simple, the Killing form is degenerate
but the stress-energy tensor can still be represented 
as a bilinear form in the currents,
\be T = \frac{1}{4}(P^+P^-+P^-P^+)+JK+\frac{\m^2}{2}K^2 \ .  \label{a3} \ee
The background metric and the antisymmetric tensor field of the corresponding
$\s$-model are:
\be ds^2 = -2 du dv - \frac{\m^2 r^2}{4} du^2 + dr^2 + r^2
d\f^2 \ , \hspace{1cm} B_{\f u} = \frac{\m r^2}{2} \ , \label{a4} \ee
and describe a four-dimensional gravitational wave.
Gravitational waves in $2+2n$ dimensions ($n \ge 1$) with similar metrics
and fluxes correspond to WZW models based on the $H_{2+2n}$ groups. 
Their central charge is $c = 2 + 2n$.

There are several reasons for being interested in this class of space-times.
First of all they are one of the few examples of curved backgrounds
where it has been possible to compute tree-level string amplitudes \cite{dk}.
Moreover they display  in a very clean context, due to their simple algebraic 
structure, all the new interesting features of the non-compact WZW models, such as
an infinite number of representations and the spectral flow.
Finally they arise as Penrose limits of space-times 
for which we know the holographic description and therefore we expect that
the analysis of the string correlators could teach us something about 
the holographic description of the pp-waves \cite{kp}. 
The most studied case is the BMN limit
of ${\cal N}=4$ SYM \cite{bmn} related to string theory in the
maximally supersymmetric wave in ten dimensions \cite{pen}. In
this example the gravitational wave is supported by a RR flux and 
the world-sheet $\s$-model has been quantized only in the light-cone
gauge \cite{metsaev}. 
On the other hand, the 
$H_4$ WZW model can be quantized in a covariant way and the three and 
four-point amplitudes have been computed using standard current algebra 
techniques, as we will discuss in the following. 
This WZW model arises as the Penrose limit of the near-horizon region of a 
collection of NS5 branes and is therefore related to the
Little String Theories \cite{gomis}.
An interesting application of our results would be the study of 
holography in the pp-wave limit of $AdS_3 \times S^3$ using 
the $H_6$ WZW model.

We start our analysis by discussing the spectrum of the $H_4$ WZW model.
For this purpose we label the states with two quantum numbers: $p$,  
the eigenvalue of $K$ which we can identify with the
momentum conjugate to the $v$ direction and $j$, 
the eigenvalue of $J$.
If we make a Fourier transformation in the light-cone directions, 
the wave equation in the pp-wave background $(3)$ reduces to 
the Schr\"odinger equation for a two-dimensional
harmonic oscillator with frequency proportional to $\m p$. As a 
consequence particles with non-zero $p$ are confined by
the gravitational wave in periodic orbits in the transverse plane
while particles with $p = 0$ do not feel the potential and 
can move freely in the transverse plane.
These three different types of motion parallel the 
three types of unitary representations of the $H_4$ algebra: 
\begin{enumerate}
\item the lowest-weight $V^+_{p,\jh}$ representations with
positive light-cone momentum $p$ and $j = \jh + n$, $n \in
\mathbb{N}$; 
\item  the highest-weight $V^-_{p,\jh}$ 
representations with negative light-cone momentum $-p$
and $j = \jh - n$, $n \in \mathbb{N}$;  
\item  the $V^0_{s,\jh}$ representations
with zero light-cone momentum and $j = \jh + n$, $n \in \mathbb{Z}$.
The positive number $s$ measures the radial momentum in the transverse plane.
\end{enumerate}

The highest-weight representations of the current algebra lead to
a ghost-free physical string state spectrum only if we restrict 
the unitary representations of the zero-mode algebra to the range 
$0 \le \m p < 1$. This bound has the same origin as the bound on the quantum
number $l$ labeling the highest-weight representations of $SL(2,\mathbb{R})_k$ 
and it is therefore not surprising that also in our case the states with $\m p \ge 1$ 
belong to the so-called spectral-flowed representations \cite{maloog1}.  
These representations are highest-weight representations 
of the isomorphic algebra $\tilde{H}_{4,w}$, $w \in \mathbb{Z}$, whose modes are related to
the original ones by
\be
\tilde{P}^{\pm}_n = P^{\pm}_{n\mp w} \ , \hspace{0.4cm}
\tilde{K}_n = K_n -i w \d_{n,0} \ ,  \hspace{0.4cm}
\tilde{J}_n = J_n  \ ,  \hspace{0.4cm} \tilde{L}_n = L_n -i w J_n \ .
\label{}
\ee
The complete spectrum of the model then contains the 
$V^+_{p,\jh}$ representations, with $\m p<1$ and
$\jh \in \mathbb{R}$, and their
spectral-flowed images $\w_w(V^+_{p,\jh})$ with $w \in \mathbb{N}$;
the $V^-_{p,\jh}$ representations, with $\m p<1$ and
$\jh \in \mathbb{R}$, and their
spectral-flowed images  $\w_{-w}(V^+_{p,\jh})$ with $w \in \mathbb{N}$; 
the $V^0_{s,\jh}$ representations, with $s \ge 0$ and $\jh \in [-1/2,1/2)$, 
and their
spectral-flowed images  $\w_{w}(V^0_{s,\jh})$ with $w \in \mathbb{Z}$.
We consider left-right symmetric combination of the representations
with the same amount of spectral flow in the two sectors.
Note that in string theory all the states with $\m p \in \mathbb{Z}$
can move freely in the transverse plane, not only the states with $p=0$.
These states are the so-called long strings \cite{sw,maloog1}.

Let us now concentrate on the highest-weight representations.
We introduce the local primary fields $\F^{a}_{q}(z,\bar
z;x,\bar{x})$ that depend on the world-sheet coordinates $z$, $\bar z$ 
and on two further variables $x$ and $\bar{x}$, that encode the
states of the left and right infinite-dimensional representations
of the left and right $H_4$ current algebras. 
In the following expressions we will often omit the
dependence on $\bar{z}$ and $\bar{x}$. Here $a$ labels
the different representations $(a \in \{+,-,0 \} )$ and $q$ stands
for the set of charges needed to completely specify a given
representation, that is $q = (p,\jh)$ when $a = \pm$ and $q=
(s,\jh)$ when $a=0$.

The introduction of the auxiliary charge variable $x$ is a very
useful tool \cite{zf}. The zero-modes of the currents can be realized as
operators acting on these variables and the OPE's can be written in 
a clear and compact form. In our case, we use the
standard realization of the $H_4$ group in terms of multiplication
and differentiation operators.
For the $V^{\pm}_{p,\jh}$ representations we introduce the fields \be
\F^{\pm}_{p,\jh}(z,x) = \sum_{n=0}^{\infty}R^{\pm}_{p,\jh;n}(z)\frac{(x
\sqrt{p})^n}{\sqrt{n!}} \ , \ee and the operators \be P_0^\pm =
\sqrt{2} p ~x \ , \hspace{0.6cm} P_0^\mp = \sqrt{2} ~\p_x \ ,
\hspace{0.6cm}  J_0 = i(\jh \pm x \p_x ) \ , \hspace{0.6cm} K_0 =
\pm i p \ . \label{} \ee The monomials $b_n =
\frac{(x\sqrt{p})^n}{\sqrt{n!}}$ form an orthonormal basis if we
define the scalar product using a gaussian measure. 
The right-moving algebra is similarly realized as an algebra of 
operators acting on the independent variable $\bar{x}$. The conformal
dimension of the primary fields is 
\be
h = \mp p \jh + \frac{\m p}{2} (1 - \m p) \ .
\ee
For the $V^0_{s,\jh}$ representations we introduce the fields \be
\F^0_{s,\jh} = \sum_{n=-\infty}^\infty R^0_{s,\jh;n}(z) x^n \ ,
\ee with $x = e^{i\a}$ and the operators \be P_0^+ = s x  \ ,
\hspace{0.6cm} P_0^- = \frac{s}{x} \ , \hspace{0.6cm} J_0 = i(\jh
+ x \p_x ) \ , \hspace{0.6cm} K_0 = 0 \ . \label{} \ee 
The conformal dimension of the primary fields is $h = s^2/2$.
The general form of the three-point couplings can now be written as 
\be \langle
\F^a_{q_1}(z_1,x_1)\F^b_{q_2}(z_2,x_2) \F^c_{q_3}(z_3,x_3)\rangle
 = \frac{{\cal C}_{abc}(q_1,q_2,q_3)
D_{abc}(x_1,x_2,x_3)}{\prod_{j>i=1}^3|z_{ij}|^{2(2h_i+2h_j-h)}}
\ , \label{} \ee where $h = \sum_{i=1}^3
h_i$. Here the ${\cal C}_{abc}(q_1,q_2,q_3)$ are the quantum structure
constants of the CFT and the functions $D_{abc}$ are the generating
functions for the Clebsch-Gordan coefficients of the left and
right $H_4$ algebras.
As an explicit example consider the fusion between two $\F^+$
representations \be [\F^+_{p_1,\jh_1}] \otimes [\F^+_{p_2,\jh_2}]
= \sum_{n=0}^\infty [\F^+_{p_1+p_2, \jh_1+\jh_2+n}] \ . \ee  
The function $D_{++-}$ is  given by 
\be D_{++-}(x_1,x_2,x_3) = \left |
e^{-x_3(p_1x_1+p_2x_2)}(x_2-x_1)^{-L} \right |^2  \ , 
\label{} \ee
and is non-zero only when $p_3=p_1+p_2$
and $L = \sum_{i=1}^3 \jh_i$ is a non-positive integer. 

The quantum structure constants ${\cal C}_{abc}(q_1,q_2,q_3)$ 
can be derived studying the factorization of the four-point amplitudes \cite{dk}, 
as we will illustrate momentarily. There are three types of couplings: \\
1) Couplings of the form $\langle V^\pm V^\pm V^\mp \rangle$, involving only $V^\pm$ representations. 
For example
\be {\cal C}_{++-}(q_1,q_2,q_3) = \frac{1}
{\G(1-\jh_1-\jh_2-\jh_3)} \left [ \frac{\g(p_3)}{\g(p_1)\g(p_2)}
\right ]^{\frac{1}{2}-\jh_1-\jh_2-\jh_3} \  , \label{cppm} \ee
where $\g(x) = \G(x)/\G(1-x)$.
The other couplings can be obtained 
from $(13)$ using the symmetry of the ${\cal C}_{abc}$ in their indexes
and the fact that ${\cal C}_{++-}={\cal C}_{--+}$, up to inverting the sign of all 
the $\jh_i$. For example
\be
{\cal C}_{+--}(q_1,q_2,q_3) =  \frac{1}
{\G(1+\jh_3+\jh_1+\jh_2)} \left [ \frac{\g(p_1)}{\g(p_2)\g(p_3)}
\right ]^{\frac{1}{2}+\jh_3+\jh_1+\jh_2} \  , \label{cpmm} \ee
with $p_3=p_1+p_2$ and $\jh_3 = -\jh_1-\jh_2+n$, $n \in \mathbb{N}$. 
We will often use a short-hand notation for these three-point couplings
writing for instance ${\cal C}_{++-}(q_1,q_2,n)$ 
to denote the coupling in equation $(13)$ with $\jh_3 = -\jh_1 - \jh_2 - n$. \\
2) Couplings of the form  $\langle V^+V^-V^0\rangle$. They are given by \be
{\cal C}_{+-0}(p,\jh_1;p,\jh_2;s,\jh_3) \ = \ 
e^{\frac{s^2}{2}[\psi(p)+\psi(1-p)-2\psi(1)]} \ , \label{cpm0} \ee
where $\psi(x) = \frac{d \ln{\G(x)}}{dx}$ is the digamma function.
We introduce a short-hand notation also for these couplings setting
${\cal C}_{+-0}(p,s) \equiv {\cal
C}_{+-0}(p,\jh_1;p,\jh_2;s,\jh_3)$. \\
3) Couplings between three $V^0$ representations. They are the same
as in flat space.

We now turn to the four-point amplitudes and explain how they can be computed
using factorization and the Knizhnik-Zamolodchikov (KZ) equations.
As it is well-known the correlation functions between the affine primary fields 
of a WZW model satisfy a system of differential equations, the 
KZ equations, as a direct consequence of the Sugawara form of the stress-energy tensor.
For a four-point amplitude the global conformal and $H_4$ symmetries 
can be used to reduce the dependence on the eight $z_i$ and $x_i$ variables to 
the dependence on only two variables, the cross-ratio $z =
\frac{z_{12}z_{34}}{z_{13}z_{24}}$ and an invariant $x$ that is
different for different types of correlators.
In this way the KZ equation becomes a partial differential equation in 
two variables.
Using the operator algebra the four-point amplitudes can be 
decomposed in a sum over intermediate representations of the affine
algebra. The functions that appear in this decomposition are called
conformal blocks and are particular solutions of the KZ equation
which satisfy suitable boundary conditions. 
The four-point amplitudes can then be reconstructed 
as a monodromy invariant combination of the conformal blocks. The requirement
of monodromy invariance arises because 
the four-point functions can be factorized in different ways
and all of them must agree in order to respect the associativity of the operator
algebra. In terms of the cross-ratio $z$, the three factorization limits are 
$z=0$, $1$ and $\infty$.

Here we discuss only the general structure of the amplitudes. Explicit expressions
can be found in \cite{dk}.
The simplest four-point functions involve three $\F^+$ and one
$\F^-$ vertex operators \be <\F^+_{p_1,\jh_1}(z_1,x_1)
\F^+_{p_2,\jh_2}(z_2,x_2)\F^+_{p_3,\jh_3}(z_3,x_3)
\F^-_{p_4,\jh_4}(z_4,x_4)> \ , \label{} \ee with $p_1+p_2+p_3 =
p_4$ and $L = \sum_{i=1}^4 \jh_i \in \mathbb{Z}$. 
These amplitudes factorize on a finite number of conformal
blocks when $L \le 0$ and vanish when $L > 0$. In the
$s$-channel the intermediate states belong to the representations
$\F^+_{p_1+p_2,\jh_1+\jh_2+n}$ with $n=0,1,...,|L|$ and the
four-point functions can be written as \be {\cal
A}(z,\bar{z},x,\bar{x}) = \sum_{n=0}^{|L|} {\cal
C}_{++-}(q_1,q_2,n) {\cal C}_{+-+}(q_3,q_4,|L|-n)|{\cal
F}_n(z,x)|^2 \ . \label{} \ee
The most interesting correlators involve two $\F^+$ and two $\F^-$
vertex operators \be <\F^+_{p_1,\jh_1}(z_1,x_1)
\F^-_{p_2,\jh_2}(z_2,x_2)\F^+_{p_3,\jh_3}(z_3,x_3)
\F^-_{p_4,\jh_4}(z_4,x_4)> \ , \label{} \ee with $p_1+p_3 =
p_2+p_4$. In this case the number of intermediate states is infinite 
and the amplitudes can be written as \be {\cal
A}(z,\bar{z},x,\bar{x}) = \sum_{n=0}^{\infty} {\cal
C}_{+--}(q_1,q_2,n) {\cal C}_{+-+}(q_3,q_4,n+|L|)|{\cal
F}_n(z,x)|^2 \ . \label{} \ee Here the conformal blocks correspond
to the representations $\F^+_{p_1-p_2,\jh_1+\jh_2-n}$ and we have
assumed $p_1 > p_2$ and $L \le 0$.
When $p_1 = p_2$ the intermediate states belong to the
$\F^0_{s,\jh_1+\jh_2}$ representations and the amplitude factorizes
as follows \be {\cal A}(z,\bar{z},x,\bar{x}) = \int_{0}^{\infty}
ds s \ {\cal C}_{+-0}(q_1,q_2,s) {\cal C}_{+-0}(q_3,q_4,s)|{\cal
F}_s(z,x)|^2 \ . \label{} \ee
Finally when $p_1+p_3 \ge 1$ we can explicitly verify by
studying the factorization of the amplitude that the intermediate states belong to
the spectral-flowed representations. The inclusion of these representations
is therefore necessary in order to define a closed operator algebra. 
The fusion rules between the spectral-flowed representations turn out to be
\be
\w_{w_1}(\F_1) \otimes \w_{w_2}(\F_2) = \w_{w_1+w_2}(\F_1 \otimes \F_2) \ ,
\label{sf2}
\ee
as first proposed in \cite{gaberdiel}.
A detailed analysis of the factorization
properties of the four-point amplitudes and of the transformations of the
conformal blocks can be found in \cite{dk}. It would be interesting to
compare the braiding and fusion matrices that exchange the different basis
of conformal blocks with the 6j-symbols of the quantum Heisenberg group.

There is another interesting aspect of the model we would like to  
briefly mention. The $H_4$ current algebra has a free field realization \cite{kk}
in term of which the primary vertex operators correspond to 
orbifold twist fields. As a consequence the $H_4$ three-point couplings
displayed before  can be compared
with the couplings computed in \cite{dfms} for the case of a rational twist.
Similarly the $H_4$ four-point correlators just discussed 
can be considered as generating functions for correlators between 
arbitrarily excited twist fields. 
Finally the three and four-point correlators of the $H_4$ WZW model
(and in principle also arbitrary $N$-point correlators) can be evaluated
using a Wakimoto representation, as recently discussed in \cite{cfs}.

The Nappi-Witten gravitational wave can be obtained as the Penrose
limit of a space of the form $\mathbb{R} \times S^3$. From the
world-sheet point of view this operation amounts to 
a contraction of the affine algebra $U(1)_{{\rm time}} \times SU(2)_k$ \cite{ors}.
In the contraction, the level $k$ is sent to infinity and
the quantum numbers of the states are scaled in such a way that
the representations of the original algebra organize themselves in representations
of the contracted one. 
Since it is by now well known how to
perform the Penrose limit in space-time, we will concentrate on
the contraction of the current algebra. Let $J^0$ be the $U(1)$
current and $J^3$, $J^\pm$ the three $SU(2)$ currents. The
contraction to the $H_4$ algebra is obtained by first changing
basis to the new currents \be K(z) = \frac{2i}{k} J^0(z) \ ,
\hspace{0.4cm} J(z) = i(J^0(z)-J^3(z)) \ , \hspace{0.4cm} P^\pm(z) =
\sqrt{\frac{2}{k}} J^\pm(z) \ , \ee and then sending the level $k$ to
infinity.

As we said, the first thing we would like to understand is how the states of
the original and of the contracted models are related. There are 
three different classes of states that are relevant in the limit.
Let us label the original states with their spin $l$ and with their 
$J^0$ and $J^3$ eigenvalues, $q$ and $m$ respectively. The quantum
numbers of the first two classes of states scale as \be  q = \pm
\frac{kp}{2} \ , \hspace{1cm} l = \frac{kp}{2} \mp \jh \ ,
\hspace{1cm} m = \pm \frac{kp}{2} - \jh \mp n \ . \label{s1} \ee These 
states in the limit are very close either to the top or to the bottom of an $SU(2)$
representation and give rise to the $V^\pm_{p,\jh}$
representations. The quantum numbers of the third class of states scale 
as $l \sim \sqrt{k}s$, $q, m \sim O(1)$ and correspond to states in
the middle of an $SU(2)$ representation. In the limit they give rise to the
$V^0_{s,\jh}$ representations. At the level of the semiclassical
wave functions the contraction amounts to a limit relating 
the Jacobi and Laguerre polynomials.

Having understood how the states are connected, we can study how
dynamical quantities such as the three and the four-point functions
behave in the limit. The $SU(2)_k$ WZW model was solved by Fateev
and Zamolodchikov in \cite{zf}. They introduced a charge variable
$x$ for $SU(2)$ and studied correlation functions 
between affine primary fields $\F_l(z,x)$.
The $SU(2)$ Ward identities completely fix the dependence  
of the three-point couplingson the $x$ variables:  
$ D(l_1,l_2,l_3) = \prod_{i<j}^3 x_{ij}^{l_{ij}}$ with 
$l_{12} = l_1+l_2-l_3$ and cyclic permutations.
The quantum structure constants of the operator algebra are \be
C^2(l_1,l_2,l_3)=\g\left(\frac{1}{N}\right)~P^2(l_1+l_2+l_3+1)\prod_{i=1}^3
{P^2(l_1+l_2+l_3-2l_i)\over\g\left({2l_i+1\over
N}\right)~P^2(2l_i)} \ , \label{fz2}\ee with \be P(n)=\prod_{m=1}^n
\g\left({m\over N}\right) \ , \hspace{1cm} P(0)=1 \ ,
\hspace{1cm} N = k+2 \ . \label{fz1}\ee
It is easy to verify that if we use the following relations between
the $H_4$ and $U(1) \times SU(2)_k$ primary fields 
\be  \Phi^{-}_{p,\jh}(x,z)=\lim_{k\to\infty}~\Phi_l\left(
\sqrt{\frac{p}{2l}}x ,z\right) \ , \hspace{1cm} 2l=kp + 2\jh \ ,
\label{}\ee \be  \Phi^{+}_{p,\jh}(x,z)=
\lim_{k\to\infty}~\left(\frac{1}{x}\sqrt{\frac{2l}{p}}
\right)^{-2l}~ \Phi_l\left(\frac{1}{x}\sqrt{\frac{2l}{p}},z\right)
\ , \hspace{0.3cm} 2l=kp-2\jh \ , \label{}\ee \be
\Phi^0_s(x,z)=\lim_{k\to\infty}~x^{-l-\jh}~\Phi_l(x,z) \ ,
\hspace{1cm} 2l=\sqrt{2k}~s \ , \label{}\ee
the  $U(1) \times SU(2)_k$ three-point couplings in equation
$(24)$ reproduce the $H_4$ couplings displayed in equations 
$(13)$, $(14)$ and $(15)$.
A discussion of the behaviour in the limit of the simplest four-point functions
and of the spectral flow can be found in \cite{dk}.

We are ultimately interested in critical string theory backgrounds of the form 
${\it C}={\it C}_{H_4} \times {\it C}_{{\rm int }} \times {\it C}_{{\rm gh}}$,
the simplest choice being ${\it C}_{{\rm int}} = \mathbb{R}^{22}$. 
The string theory amplitudes are given by the CFT correlators
integrated over the moduli space of the corresponding surface with
punctures. For the four-point functions this means that we have to
integrate over the cross-ratio $z$. The most interesting feature of
the string amplitudes in this pp-wave background 
is that the propagation of long string states in the intermediate 
channels gives rise to a logarithmic branch cut.

The last aspect we would like to discuss is the flat space limit
of the pp-wave metric \cite{dk}. The value of the parameter $\m$  in equation
$(3)$ can be changed performing a boost $u \rightarrow \l u$, $v \rightarrow
v/\l$. There are however two interesting limits to consider:
$\m \rightarrow 0$ and $\m \rightarrow \infty$.
In the limit $\m \rightarrow 0$ the metric reduces to the metric of 
flat Minkowski space and the current algebra to the algebra of four 
free bosons. We want to emphasize that this limit can be considered  
as a contraction of the algebra $H_4$ to $U(1)^4$ and can be discussed 
in exactly the same way as we did for the contraction from $U(1) \times SU(2)$ 
to $H_4$.  Note that these contractions are singular limits and in particular 
they change the asymptotic structure of the space-time. 
At the level of the semiclassical wave-functions the contraction $\m \rightarrow 0$
amounts to a limit relating the Laguerre polynomial and the Bessel
functions.

In the  $\m \rightarrow 0$ limit we obtain flat space as suggested by the
classical intuition: the potential
flattens and the states trapped by the wave describe larger and
larger orbits until they become free. All the states in flat space
arise from states in the highest-weight representations of the $H_4$
WZW model, while the spectral-flowed representations are
scaled out of the spectrum.
We would like to point out that also the limit $ \m \rightarrow
\infty$ leads to flat space even though the states that survive in
the limit are markedly different. Indeed in this case  the
flat space vertex operators arise from the spectral-flowed continuous 
representations, while all the other representations with $\m p
\notin \mathbb{Z}$ are so strongly trapped by the potential that
disappear from the spectrum. This is a typical stringy mechanism and 
it is similar under some respect to the large and small radius limit
of a compactified boson.

\ack

It is a pleasure to thank Elias Kiritsis for a very enjoyable 
collaboration and the organizers of the RTN Workshop 
in Copenhagen for giving me the opportunity to present our results.
The author is supported by an European Commission Marie Curie Individual 
Fellowship, contract HPMF-CT-2002-01908. The work reported here
was also partially supported by RTN contracts HPRN-CT-2000-00122 
and  HPRN-CT-2000-00131.

\end{document}